\documentclass{JHEP3}

\usepackage{latexsym}
\usepackage{amsfonts}

\newcommand{\bm}[1]{\mbox{\boldmath $#1$}}

\def\xiv{\vec \xi }

\def\lie{{\pounds}}

\def\S{\Sigma}
\def\di{{\rm div}}

\def\be{\begin{equation}}
\def\ee{\end{equation}}
\def\bea{\begin{eqnarray}}
\def\eea{\end{eqnarray}}
\def\bean{\begin{eqnarray*}}
\def\eean{\end{eqnarray*}}

\title{On the existence of horizons in spacetimes with vanishing curvature 
invariants}

\author{Jos\'e M M Senovilla\thanks{Financial support from
grants BFM2000-0018 (Spanish CICyT) and 
9/UPV 00172.310-14456/2002 (University of the Basque 
Country) is acknowledged.}\\
Departamento de F\'{\i}sica Te\'orica, Universidad del Pa\'{\i}s Vasco,
Apartado 644, 48080 Bilbao, Spain \\
E-mail: \email{wtpmasej@lg.ehu.es}}

\abstract{
A direct very simple proof that there can be no closed trapped 
surfaces (ergo no black hole regions) in spacetimes with all 
curvature  scalar invariants vanishing is given. 
Explicit examples of the recently 
introduced ``dynamical horizons'' which nevertheless do not enclose any trapped 
region are presented too.}

\preprint{}

\keywords{bhs, ctg, sts, dag}

\begin{document}

There is a renewed interest on (generalized) pp-wave 
spacetimes\footnote{See e.g. \cite{Exact} and references therein 
for the standard
definition and properties of pp-waves in General Relativity, and 
\cite{H,HR1} for generalizations to arbitrary 
dimensions.}. This is due to the fact that all scalar 
invariants constructable from the  Riemann tensor and its derivatives 
vanish in pp-wave spacetimes, which implies that they are {\em exact} solutions
of the full non-linear classical string theory \cite{alpha}. 
Moreover, it is known 
since long ago that every spacetime can be ``approximated'', nearby a 
null geodesic, to a plane wave (a particular case of pp-waves, see e.g.
\cite{Exact,HE,MTW}), which is called the Penrose limit of the spacetime on 
that geodesic \cite{P}. This has relevant consequences and 
applications in the context of the AdS/CFT correspondence, 
see e.g. \cite{BMN,limit} and references therein. The above led to the 
analysis of the  pp-wave conformal boundary, linked via AdS/CFT to the 
associated conformal theory, with the remarkable result that pp-wave 
spacetimes seem to have a single null line as conformal boundary 
\cite{null-b,HR1}. 

More generally, spacetimes with all scalar 
invariants vanishing (VSI from now on) share the above mentioned property of 
being exact solutions of classical string theory \cite{alpha,C}. Thus, they are 
claimed to give some insight into an acceptable theory of quantum 
gravity. The question has been raised of whether these exact 
solutions of string theory can describe, or contain, black-hole 
regions, something which would be expected to provide deeper or 
clearer hints on the path to the sought quantum theory of gravitational 
fields. 
As a matter of fact, using techniques of geodesic connectivity, one 
can demonstrate that (some) pp-waves can have no ``event horizon'' \cite{HR2} 
---in the sense that every point in the spacetime can be joined to infinity
by means of a causal future-directed curve---, see also \cite{FS}. 
However, the question remains open for general VSI spacetimes, 
as has been recently pointed out in \cite{HR3}.

In this short note, I address the issue and give a direct very simple proof of 
the complete absence of closed trapped surfaces (and more generally of closed 
trapped submanifolds of any co-dimension) in VSI spacetimes. Interestingly, 
the proof relies on 
recently developed (i) concepts generalizing Killing vectors
---Kerr-Schild symmetries \cite{KS}---and (ii)
arguments on the interplay between symmetries and trapped surfaces \cite{MS}; 
this is indication of potential applications of both (i) and (ii) 
to general theories based on Lorentzian 
geometry. I intend to call attention to these works by means of the
particular application treated herein. The proof will be essentially 
geometric and no field equations or any other conditions are assumed. 
As a by-product, closed {\em marginally} trapped surfaces will come 
out to be ubiquitous in VSI spacetimes,
and thereby explicit examples of {\em dynamical horizons} in the 
sense of \cite{As} will be exhibited, showing that this kind of horizons do 
not enclose trapped regions or black holes in general. 

Consider the VSI spacetime (signature --,+,\dots ,+), 
in arbitrary dimension $D$, given in local coordinates $\{u,v,x^i\}$ 
($i,j,\dots =1,\dots ,D-2$) by the line-element \cite{C}
\be
ds^2=-2du(dv+Hdu+W_{i}dx^i)+g_{ij}dx^idx^{j} \label{et}
\ee
where $H=H(u,v,x^i)$ is arbitrary, the functions 
$g_{ij}=g_{ji}=g_{ij}(u,x^{k})$ are independent of $v$, and 
$W_{i}$ are linear on $v$, that is 
$$
W_{i}=vF'_{i}+Z_{i}(u,x^{k})
$$
the $v$-coefficient $F'_{i}$ depending only on the 
corresponding $x^i$ (primes stand for derivatives with respect to the 
argument)\footnote{By using 
the remaining coordinate freedom \cite{C,Exact}, one can rewrite the above 
such that $W_{1}=vF'_{1}(x^1)+Z_{1}(u,x^{k})$ with $F'_{1}(x^1)=dF_{1}(x^1)/dx^1$, 
and $W_{j}=Z_{j}(u,x^{k})$ for all $j\neq 1$. However, this is not 
necessary here. Analogously, 
one could allow for an arbitrary function in front of $dv$ in 
(\ref{et}).}. There is a 
preferred null vector field $\vec\ell =\partial_{v}$ in these spacetimes, 
characterized by being a gradient (hence geodesic), shear-free and 
expansion-free. Its covariant form is $\bm{\ell}=-du$. Therefore, 
(\ref{et}) belongs to the so-called Kundt class (generalized to $D$ 
dimensions) \cite{Ku,Exact}. The pp-waves are characterized by admitting 
the existence of a 
covariantly constant null vector field $\vec\ell$, whence they are 
included in (\ref{et}) when $F'_{i}=0$ and $H_{,v}=0$ ---a comma 
indicates partial derivative---. 

In general, though, the null vector field 
$\xiv\equiv \exp\left(-\sum_{i}F_{i}\right)\vec\ell$
still satisfies a quite interesting relation, namely
$$
\lie_{\xiv}\, g =-2 F^{-1} H_{,v}\, \bm{\ell}\otimes\bm{\ell}
$$
where $\lie_{\xiv}$ stands for the Lie derivative with respect 
to $\xiv$ and $F>0$ is a 
shorthand for $\exp\left(\sum_{i}F_{i}\right)$.
This is the differential condition defining the so-called {\em 
Kerr-Schild vector fields} \cite{KS}, which are the generators of one-parameter 
groups of Kerr-Schild transformations. A study of these 
infinitesimal symmetries, their properties, and some applications can 
be found in \cite{KS}. Even more importantly, the crucial point in 
what follows is that, actually, there is an {\em infinite number} of Kerr-Schild 
vector fields, all proportional to $\vec\ell$, depending on an 
arbitrary function of $u$. This follows straightforwardly from 
Lemma 3.2 and Theorem 2 in \cite{KS}, or alternatively by a direct 
calculation:
\be
\lie_{\xiv_{f}}\, g = -2 F^{-1}\left[f(u)H_{,v}+f'(u)\right]
\bm{\ell}\otimes\bm{\ell} \label{defo}
\ee
for all vector fields
\be
\xiv_{f}\equiv f(u) \, F^{-1}\,  \vec\ell \label{ksvfs}
\ee
$f(u)$ being an {\em arbitrary} function---thus, the Lie algebra of 
Kerr-Schild vector fields with respect to $\vec\ell$ is 
infinite-dimensional---. Let us remark that the above statement holds for 
pp-wave spacetimes too (including plane waves and even flat 
spacetime): of course, there are not Killing vectors depending on 
arbitrary functions, but there certainly are Kerr-Schild vector fields 
with that property, and they happen to include, for a choice of the 
arbitrary function $f(u)$ (up to a constant of proportionality),
a Killing vector.

The arguments of \cite{MS} can now be applied. Let $S$ be any 
spacelike submanifold of any dimension $d$, and let 
$\{\vec{e}_{A}\}$ ($A,B,\dots =1,\dots ,d$) denote a set of 
$d$ linearly independent tangent vector fields to $S$ on $V$. Let 
$\gamma_{AB}=g|_{S}(\vec{e}_{A},\vec{e}_{B})$ be the first 
fundamental form inherited by $S$ and $\overline\nabla$ its canonical 
connection. In \cite{MS} we proved that, for {\em any} vector field $\xiv$,
\be
\frac{1}{2}
\gamma^{AB}\lie_{\xiv} g|_S (\vec{e}_{A},\vec{e}_{B})=
\overline\nabla_{C}\overline\xi^{C}+ (\xi_{\mu} 
{\cal H}^{\mu})|_{S}\label{paso}
\ee
where for all $\vec v$, $\overline{v}_{C}=v_{\mu}|_{S}\, e^{\mu}_{C}$,
and $\vec{{\cal H}}$ denotes the mean curvature vector of $S$ (e.g. \cite{KH,Kr,S2,MS}). 
Future trapped submanifolds are characterized by having
$\vec{{\cal H}}$ pointing to the future all over $S$, and similarly for past 
trapped. The trapping is proper, near, or marginal according to 
whether $\vec{{\cal H}}$ is timelike, non-spacelike, or null and 
non-vanishing all over $S$ \cite{MS,HE,KH}. Therefore, if $S$ is 
trapped and $\xiv$ is future (or past) pointing all over $S$, the 
second term on the righthand side of (\ref{paso}) cannot change sign.

Suppose, then, that $S$ is (marginally, nearly) future trapped and  
closed (i.e. compact without boundary; this is the case of interest 
for black-hole regions). Replacing the Kerr-Schild symmetries 
(\ref{ksvfs}) for $\xiv$ in (\ref{paso}) ---with $f(u)$ positive, say,
so that $\xiv_{f}$ are future pointing---, integrating the 
resulting relation on $S$, using (\ref{defo}) and Gauss' theorem, one 
arrives at
\be
-\int_{S} \left(\gamma^{AB}\overline{\ell}_{A}\overline{\ell}_{B}\right)
F^{-1}\left[f(u)H_{,v}+f'(u)\right]\bm{\eta}_{S} = 
\int_{S} F^{-1} f(u)(\ell_{\mu} {\cal H}^{\mu})\, \bm{\eta}_{S} \leq 0 \label{main}
\ee
where $\bm{\eta}_{S}$ is the canonical $d$-volume element on $S$. 
Recalling that $f(u)$ is an arbitrary function, this 
inequality clearly leads to a contradiction in general. To prove it rigourously, 
notice that $\gamma^{AB}\overline{\ell}_{A}\overline{\ell}_{B}\geq 0$, 
and this vanishes if and only if 
$\overline{\ell}_{A}=\ell_{\mu}|_{S}e^{\mu}_{A}=0$. If 
$\overline{\ell}_{A}$ were non-zero at some points of $S$, one could always 
choose $f(u)$ such that $\left(f(u) H_{,v}+f'(u)\right)|_{S}\leq 0$ 
which would give the ``wrong'' 
sign for the left integral in (\ref{main}). Indeed, as $S$ is 
compact, $H_{,v}$ will reach its maximum $M$ on $S$, 
hence $H_{,v}|_{S}\leq M$, so that it 
would be enough to choose a positive $f$ with $f'/f\leq -M$ (for instance, 
$f(u)=e^{-Mu}$ would do). The only possibility is therefore that 
$\overline{\ell}_{A}=0$. Then the 
right integral of (\ref{main}) must vanish as well, whence 
$\ell_{\mu} {\cal H}^{\mu}=0$. As $\vec{{\cal H}}$ and $\vec \ell$ are causal this 
implies that in fact they must be proportional to each other. In 
summary

{\em The VSI spacetimes (\ref{et}) do not admit closed trapped or nearly trapped 
submanifolds of any dimension. And any closed marginally trapped submanifold 
must have $\vec{{\cal H}}\propto\vec\ell$ and be contained in one of the null 
hypersurfaces $u=$const.\ orthogonal to the null vector field $\vec\ell$.}

{\bf Remark:} Observe that this result is valid for {\em completely} general 
pp-waves as well, as was proved already in \cite{MS}.

The meaning of the general result, when applied to submanifolds of 
co-dimension 2, is that there cannot be event horizons for 
asymptotically flat cases, if any. More generally, the absence of closed 
trapped surfaces implies that there are no ``trapping horizons'' in the sense of
\cite{Hay}. These are horizons defined locally, without any reference 
to infinity, describing the boundary of black (or white) holes, see 
\cite{Hay}. Thus, there cannot be black hole regions in the VSI spacetimes. 

The above leads also to the absence of {\em apparent 
horizons}---see e.g. \cite{HE,Wald} for the 
asymptotically flat case and \cite{KH} for the general case---, which is 
defined roughly as boundary of the set of closed trapped surfaces in 
the spacetime, and itself defines (under certain assumptions of 
continuity) closed marginally trapped surfaces \cite{KH,Wald}. 
Let us remark, however, that the recently proposed definition 
of ``dynamical horizons'' \cite{As}, trying to improve that of 
trapping horizons, does not capture the absence of black holes in the 
VSI spacetimes. As a matter of fact, one can find many examples of 
dynamical horizons in VSI spacetimes. To see this, observe that {\em 
all} surfaces $S_{u,v}$ of co-dimension 2 given by constant values of the 
coordinates $u$ and $v$ are marginally trapped, as follows from a 
trivial calculation using for instance the formulas presented in 
\cite{S2} applied to (\ref{et}):
$$
\vec{{\cal H}}=\left.\left(U_{,u}-\di \vec W\right)\right|_{S_{u,v}}\, 
\vec\ell \, .
$$
Here $U=\log\sqrt{\det g_{ij}}$, $\bm{W}=W_{j}dx^j$, and $\di$ is the 
divergence on each $S_{u,v}$. Hence, the 
mean curvature vector of these surfaces is always null and the 
expansion $\theta_{\ell}$ corresponding to the null normal 
$\vec\ell$ vanishes. Take then, for 
example, the hypersurfaces $\S:\, \, v=h(u)$. These hypersurfaces are 
spacelike if $(2H+\hat{g}^{ij}W_{i}W_{j})|_{\S}<2\dot{h}$, 
where $\hat{g}^{ij}$ is the inverse matrix of $g_{ij}$, 
and they are foliated by the
marginally trapped surfaces $S_{u,v}$. Choosing these to be closed, 
which is obviously possible, all conditions in the definition of 
dynamical horizons hold for $\S$ by choosing 
$\left.\left(U_{,u}-\di \vec W\right)\right|_{\S}$ to be strictly positive 
(this is minus the ``inwards'' expansion). Note that this is valid even for 
the simpler pp-waves\footnote{Nevertheless, $\S$ are not trapping horizons, 
as the derivative of the vanishing expansion along the 
inward null direction vanishes. So, trapping horizons seem to be better 
adapted to rule out cases like the VSI spacetimes.}. 

There remains the question of validity of the local coordinates, 
or whether or not the line-elements (\ref{et}) are 
extensible\footnote{Note, for instance, that the exterior $r>2m$ region 
of Schwarzschild solution is static, so that 
similar arguments apply \cite{MS} and the absence of closed trapped surfaces is 
obvious. The trapped surfaces of the Schwarzschild black hole appear 
in the added region after the spacetime has been extended.}, see e.g. 
\cite{HE,S1}. In the case 
of usual pp-waves
the coordinates are globally defined, as is known, and the spacetime 
is geodesically complete. See \cite{FS} for a recent 
discussion of when this can be generalized to some subcases of 
(\ref{et}) with $W_{i}=0=g_{ij,v}$. In the general case (\ref{et}), problems 
may arise depending on the explicit form of the functions $H,W_{i}$ 
and $g_{ij}$. The vanishing of $\det g_{ij}$ at some points will 
usually indicate either a curvature singularity or a problem with the 
completeness of the spacelike submanifolds spanned by the $\{x^i\}$, 
and occasionally extensibility of causal geodesics. Similarly, if  
$F'_{i}$ or $H$ reach unbounded values then a curvature singularity appears 
generically. There may be, though, particular cases in which 
the spacetime is extensible. 
However, if the spacetime were extensible {\em keeping the 
VSI form} (\ref{et}), then a similar argument would apply to the added region. 
While if the VSI property is lost through the extension, then this is no longer 
an exact solution of the classical string theory and the question 
loses its main interest.


\acknowledgments
I thank Veronika Hubeny for bringing this problem to my attention, and 
Marc Mars for comments and corrections.

\end{document}